\documentclass[9pt,conference, a4paper]{IEEEtran} 

\IEEEoverridecommandlockouts
\usepackage{cite}
\usepackage{amsmath,amssymb,amsfonts}
\usepackage{algorithmic}
\usepackage{graphicx}
\usepackage{textcomp}
\usepackage{eso-pic}
\usepackage{booktabs}      
\usepackage{multirow}      
\usepackage{threeparttable} 
\usepackage{arydshln}      
\usepackage{bm}            
\usepackage{enumitem}      

\renewcommand{\footnoterule}{%
  \kern -3pt
  \hrule width 0.4\columnwidth height 0.4pt
  \kern 2.6pt
}

\usepackage{xcolor}
\def\BibTeX{{\rm B\kern-.05em{\sc i\kern-.025em b}\kern-.08em
    T\kern-.1667em\lower.7ex\hbox{E}\kern-.125emX}}
\begin{document}

\title{Multi-Teacher Distillation for Cross-Domain Streaming Electrolaryngeal Speech Encoding}


\author{
  \IEEEauthorblockN{
    Benedikt Mayrhofer$^{\star \ddagger}$, 
    Enrique Orozco Olivares$^{\star \ddagger}$, 
    Franz Pernkopf$^{\star}$, 
    Philipp Aichinger$^{\dagger \ddagger}$,
    Martin Hagmüller$^{\star \ddagger}$
  }
  \vspace{0.3em}
  \IEEEauthorblockA{
    $^{\star}$Signal Processing and Speech Communication Laboratory, Graz University of Technology\\
    $^{\dagger}$Dept. of Otolaryngology, Head \& Neck Surgery, Div. of Phoniatrics-Logopedics, Speech \& Hearing Sci. Lab, Med. Univ. of Vienna \\
    $^{\ddagger}$Comprehensive Center for AI in Medicine, Medical University of Vienna
  }
}

\maketitle

\AddToShipoutPictureFG*{%
  \AtPageLowerLeft{%
    \hspace{1.9cm}\raisebox{1.3cm}{%
      \parbox{0.85\textwidth}{\footnotesize\noindent
      \copyright~2026 IEEE. Personal use of this material is permitted. Permission from IEEE must be obtained for all other uses, in any current or future media, including reprinting/republishing this material for advertising or promotional purposes, creating new collective works, for resale or redistribution to servers or lists, or reuse of any copyrighted component of this work in other works.
      }%
    }%
  }%
}

\begin{abstract}
Self-supervised learning (SSL) has improved speech representations, yet performance degrades in pathological domains such as electrolaryngeal (EL) speech, and the computational footprint of SSL models limits their applicability in real-time, on-device deployment. We propose a multi-teacher knowledge distillation framework to train a lightweight, streaming content encoder that generalizes across healthy (HE) and EL speech. Two teachers are distilled progressively: a frozen SSL model providing discrete phonetic cluster targets from HE speech, and an EL-fine-tuned speech recognition model supplying continuous bottleneck feature targets. Evaluated via downstream speech recognition, our approach reduces the EL word error rate to 21.2\%, compared to 39.3\% for the strongest zero-shot SSL baseline. Among causal convolutional, Transformer, Conformer, and Mamba-based student architectures, a Mel-Conformer achieves the best combination of EL accuracy and computational efficiency. The final encoder contains 21.9\,M parameters and runs at a real-time factor of 0.30 under ONNX Runtime on a single CPU core.
\end{abstract}

\begin{IEEEkeywords}
electrolaryngeal speech, knowledge distillation, speech encoder, self-supervised learning, real-time inference
\end{IEEEkeywords}

\section{Introduction}
\label{sec:intro}

Patients who undergo total laryngectomy, often due to advanced laryngeal cancer, suffer the removal of the larynx, including the vocal folds, losing their biological mechanism for fundamental frequency generation. Among the limited rehabilitation options available, many patients rely on an electrolarynx, a medical device held against the neck that mechanically generates artificial excitation signals to be articulated by the remaining vocal tract~\cite{fuchs2016bionic}. Despite its utility, EL speech is characterized by a robotic, monotonous pitch and is often heavily masked by the radiated mechanical noise of the device~\cite{YaogenEL2023}. This results in speech that is perceived as highly unnatural and suffers from limited intelligibility and prosody~\cite{Lester2024, mayrhofer2025maveba_vc}.

Modern voice conversion (VC) techniques offer a promising solution for enhancing EL speech by mapping these atypical signals to HE target voices~\cite{YaogenEL2023, Lester2024, mayrhofer2025maveba_vc, MaDing2025, mayrhofericassp}. State-of-the-art VC frameworks increasingly leverage self-supervised learning (SSL) representations, such as those from HuBERT~\cite{huber2021} or WavLM~\cite{wavlm2022}, to extract linguistic content disentangled from speaker identity~\cite{FreeVC2023, QuickVC2023, AdaptVCKim2025}. However, these foundation models are primarily pretrained on massive datasets of HE speech, leading to a great domain mismatch when applied zero-shot to pathological speech, causing misalignment and poor intelligibility in downstream applications~\cite{huber2021, wavlm2022, Chungw2v, Violeta2024}. A natural remedy is to fine-tune the SSL model on EL data. However, this is prone to catastrophic forgetting~\cite{Lester2024, vandereeckt2023using}, degrading the HE-domain representations that a VC system must preserve to synthesize natural target speech. Consequently, a single encoder that generalizes across both domains without sacrificing either cannot be obtained by sequential adaptation of existing foundation models alone.

Another challenge for on-device deployment is the computational demand of these SSL models~\cite{kanagawa24b_interspeech}. High-performing foundation models typically use Transformer encoders exceeding 300 million parameters, which require an entire utterance context to generate accurate representations~\cite{huber2021, wavlm2022, Chungw2v}. For assistive technologies to be effective in real-time communication, like live conversation or video conferencing, systems must support low-latency streaming capabilities~\cite{ConanBowTie2024, Liu2026VoxtralR}. SSL models are not suited for deployment on commodity hardware in real-time scenarios, as their size and non-causal processing preclude low-latency streaming~\cite{ma2024lowlatency, kudlur2026moonshinev2}.

While lightweight convolutional encoders have been adopted in recent
real-time VC systems~\cite{yang2024streamvc, liu2025rtvc, guo2025synthvc},
these architectures were not originally designed for cross-domain pathological
speech transfer. Adapting such pipelines to EL-to-HE VC yields promising
results but leaves notable intelligibility gaps~\cite{mayrhofericassp},
motivating a dedicated domain-robust content encoder.

In this paper, we propose a lightweight, low-latency streaming-compatible encoder trained via a multi-teacher distillation framework~\cite{yang2023knowledge, Wei2025MultiDistillationFS}, to address the challenges of domain mismatch and real-time deployment constraints. Our approach bridges the domain gap by progressively distilling knowledge from two distinct teachers: (1) a ``healthy speech teacher'' using a frozen SSL model to provide robust, discrete phonetic targets for linguistic content preservation~\cite{yang2024streamvc, niekerk2022comparison, quamer2026tvtsyn}, and (2) a ``pathological speech teacher'' based on an EL-fine-tuned automatic speech recognition (ASR) model that provides continuous bottleneck features (BNFs)~\cite{Lester2024, kashkin2023hifivc}. 

We evaluate student architectures spanning causal convolutional~\cite{zeghidour2022soundstream, wu2023audiodec}, Transformer~\cite{vaswani2017attention}, Conformer~\cite{gulati20_interspeech}, and Mamba~\cite{gu2024mamba} sequence-modeling cores. We measure performance through character and word error rate (CER/WER) metrics using downstream ASR probing on HE and EL speech datasets to assess linguistic fidelity. Furthermore, we benchmark the computational efficiency of each student backbone on a single-core CPU to ensure feasibility for low-latency deployment. Our findings show that multi-teacher distillation is both necessary and
sufficient for cross-domain EL generalization, with all  four causal students outperforming zero-shot SSL baselines on EL speech and the best student
achieving an EL WER of 21.2\%, versus 39.3\% for the strongest SSL baseline, at an ONNX Runtime real-time factor (RTF) of 0.30.

\textbf{Technical Contributions:} (1) A novel cross-domain multi-teacher
distillation framework combining discrete phonetic targets from a frozen
HE SSL model with continuous BNFs from an EL-fine-tuned
ASR model, explicitly bridging the domain gap while
preserving HE representations. (2) An ablation study evaluating 7 SSL teacher
configurations and 6 streaming-compatible student backbones. (3) Quantitative validation showing significant EL WER reductions with competitive HE performance retained across all student architectures, while achieving real-time streaming speeds on single-core CPU hardware.

 
\section{Resources}
\label{sec:resources}

To train the multi-teacher distillation framework and ensure robust generalization across both typical and atypical voice domains, we compile a large-scale HE corpus alongside a low-resource EL dataset, summarized in Table~\ref{tab:datasets}. All speech data used in this work is German, sourced from Austria and Germany.

\textbf{Healthy Speech Corpus:} 
The HE speech dataset comprises approximately 2,000 hours of audio. This corpus contains several open-source datasets, including Common Voice v22~\cite{ardila2019commonvoice}, Multilingual LibriSpeech (MLS) German~\cite{pratap2020mls}, VoxPopuli~\cite{wang-etal-2021-voxpopuli}, HUI-Audio-Corpus-German~\cite{puchtler2021huiaudio}, CML-TTS~\cite{oliveira2023cml}, a subset of Emilia German~\cite{emilia2024} (filtered to exclude samples with a Deep Noise Suppression Mean Opinion Score (DNSMOS)~\cite{cumlin2024dnsmos} below 3.35), and GRASS~\cite{schuppler2014grass}, supplemented by HE parallel speakers from the Electrolaryngeal and Healthy (ELHE) Speech Corpus~\cite{Fuchs2024ELHE}. All audio recordings were resampled to 16\,kHz and converted to mono-channel format.

\textbf{Electrolaryngeal Speech Corpus:}
Acquiring EL speech data remains challenging due to patient scarcity and clinical privacy constraints. We collected a 10\,hour EL dataset from three distinct sources. The primary foundation is the pathological subset of the aforementioned ELHE parallel corpus~\cite{Fuchs2024ELHE}, featuring approximately 20 EL speakers, comprising up to 500 utterances recorded per speaker. We supplemented this with our own in-studio recordings from an ongoing clinical collection (IRB approval, Medical University of Vienna, No.\ 1790/2025; currently comprising 4 EL speakers, yielding roughly 350 utterances each). Finally, we collected 6 additional EL speakers sourced from publicly available media and YouTube videos. 

\begin{table}[htpb]
\centering
\caption{Summary of training and evaluation corpora.}
\label{tab:datasets}
\setlength{\tabcolsep}{7.0pt}
\begin{tabular}{@{}lllc@{}}
\toprule
\textbf{Domain} & \textbf{Dataset} & \textbf{Region} & \textbf{Size} \\ \midrule
\multirow{8}{*}{\textbf{HE}} 
 & Common Voice v22~\cite{ardila2019commonvoice} & Germany & \multirow{8}{*}{$\sim$\,2,000\,h} \\
 & MLS German~\cite{pratap2020mls} & Germany & \\
 & VoxPopuli (DE)~\cite{wang-etal-2021-voxpopuli} & Germany & \\
 & HUI Dataset~\cite{puchtler2021huiaudio} & Germany & \\
 & CML-TTS~\cite{oliveira2023cml} & Germany & \\
 & Emilia (Subset)~\cite{emilia2024} & Germany & \\
 & GRASS~\cite{schuppler2014grass} & Austria & \\
 & ELHE (HE part)~\cite{Fuchs2024ELHE} & Austria & \\ \midrule
\multirow{3}{*}{\textbf{EL}} 
 & ELHE (EL part)~\cite{Fuchs2024ELHE} & Austria & \multirow{3}{*}{$\sim$\,10\,h} \\
 & Own in-studio Collection (Ongoing) & Austria & \\
 & Web/YouTube Sources & Mixed & \\ \bottomrule
\end{tabular}%
\vspace{-0.30cm}
\end{table}

\section{Methodology}
\label{sec:method}

\subsection{Multi-Teacher Distillation Framework}
\label{sec:method_multiteacher}

\begin{figure}[t]
  \centering
    \includegraphics[width=1.0\linewidth]{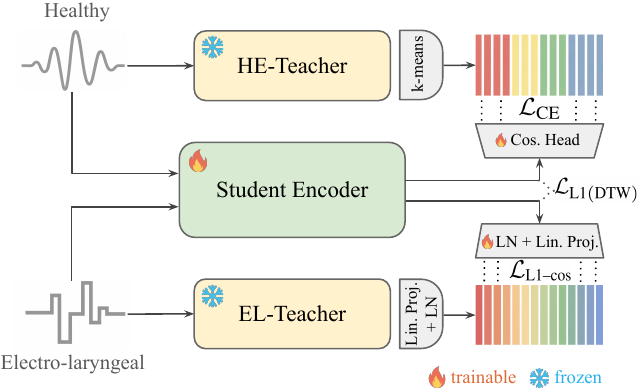}
    \vspace{-0.5cm}
  \caption{Multi-teacher distillation training paradigm.
$\mathcal{L}_{\text{CE}}$: Cross-entropy on HE discrete targets;
$\mathcal{L}_{\text{L1}\text{--}\text{cos}}$: EL bottleneck regression;
$\mathcal{L}_{\text{L1}(\text{DTW})}$: DTW-guided EL--HE alignment.}
  \label{fig:multiteacher}
  \vspace{-0.3cm}
\end{figure}

To bridge the domain gap between HE and EL speech, we propose a multi-teacher knowledge distillation paradigm that optimizes a lightweight, streaming student encoder via a 3-phase progressive training approach. Fig.~\ref{fig:multiteacher} illustrates the overall framework, in which the student encoder is supervised by two teacher branches. Training proceeds incrementally: each phase introduces one additional objective, which then remains active alongside all previously introduced objectives (Eq.~\ref{eq:total_loss}), progressively extending the student's supervision from the HE domain to joint HE--EL training and finally to explicit cross-domain alignment.

\textbf{Healthy Speech Teacher (Phase 1):} 
The primary objective of the HE teacher is to teach the student the linguistic content of the speech signal. We use a frozen, pre-trained SSL model as the HE teacher (explored in Section~\ref{sec:experiments}). Following the SoftVC approach~\cite{niekerk2022comparison}, a k-means 
clustering algorithm ($K{=}100$) is fit to the continuous HE SSL 
representations to yield discrete phonetic content labels. These representations are taken from mHuBERT-147 layer~10 (within the best-performing layer range 9--10; Fig.~\ref{fig:teacherablation}), and the k-means is fit on the full $\sim$\,2{,}000\,h HE corpus of Table~\ref{tab:datasets}. To align with this target space, the student's output is passed through a cosine similarity based classifier head comprising $K{=}100$ learned centroid embeddings. Both the 
student output and the centroids are $\ell_2$-normalized. The resulting cosine logits are scaled by a fixed temperature $\tau{=}0.1$
before computing a standard cross-entropy loss, $\mathcal{L}_{\text{CE}}$. During this phase, the student is trained exclusively with the HE dataset for approximately 300k steps (batch size 32).

\textbf{Pathological Speech Teacher (Phase 2):} 
Phase 2 transitions to a joint multi-teacher setup with mixed batches of 32 HE and 64 EL audio chunks per step, introducing the EL teacher objective only after Phase 1 convergence: given the large HE/EL size disparity (Table~\ref{tab:datasets}), joint training from initialization would risk overfitting to the small EL set before HE representations stabilize. As the EL data is too sparse for reliable k-means clustering, we instead employ continuous feature regression. The EL teacher (mHuBERT-147-BNF-CTC) is an mHuBERT-147 encoder appended with a 100-dimensional linear projection and a Connectionist Temporal Classification (CTC) head~\cite{graves2006connectionist}, fine-tuned for ASR on the combined HE and EL data. Training the recognizer jointly on both HE and EL data is preferred over
EL-only fine-tuning to limit HE representation degradation, a risk well-documented
when adapting SSL models to out-of-domain speech~\cite{Lester2024, vandereeckt2023using}.
The 100-dimensional bottleneck constrains the output to retain linguistic 
content while discarding speaker identity and non-linguistic 
acoustic variability. The student learns to approximate these continuous BNFs via $\mathcal{L}_{\text{L1}\text{--}\text{cos}}$ introduced in~\cite{Wei2025MultiDistillationFS}:

\begin{equation}
    \mathcal{L}_{\text{L1}\text{--}\text{cos}} = \frac{1}{D} \lVert h^S - h^T \rVert_1 - \log \sigma \big( \text{cossim}(h^S, h^T) \big)
    \label{eq:l1_cos}
\end{equation}

where $h^S$ and $h^T$ denote the projected representations of the student and the EL teacher, $D{=}100$ is the feature dimension, $\sigma$ represents the sigmoid activation function, and $\text{cossim}(\cdot, \cdot)$ calculates the cosine similarity between teacher and student features.

\textbf{Cross-Domain Alignment (Phase 3):} 
While the dual-teacher setup improves WER, experiments showed that 
the student maps HE and EL speech to well-separated, domain-distinct regions (Figure~\ref{fig:umap}) within the latent space. Consequently, Phase 3 is introduced with an explicit content alignment 
loss, once the student can already produce phonetically structured representations for both domains. Using parallel EL-HE pairs, we extract representations from a Whisper-large-v3~\cite{radford2023robust} encoder fine-tuned jointly on HE and EL speech to compute a Dynamic Time Warping (DTW) path, adapted from~\cite{mayrhofericassp}. No audio is resampled or time-stretched, the DTW path serves solely as a 
correspondence index identifying which EL-HE frame pairs enter the L1 penalty. 
$\mathcal{L}_{\text{L1}(\text{DTW})}$ is then computed strictly along this path, forcing the student 
to project acoustically disparate but phonetically equivalent frames from both domains closer 
together. The HE features entering $\mathcal{L}_{\text{L1}(\text{DTW})}$ are detached from the 
computation graph, serving as a fixed anchor. Gradients flow only through the 
EL branch to avoid corrupting the stable HE representations.

\textbf{Data Augmentation:}
All training audio samples are cropped to random 3\,s chunks at random 
positions. To improve robustness to acoustic variability, we apply the following stochastic online augmentation\footnote{https://github.com/iver56/torch-audiomentations} 
independently per audio chunk across all training phases, where $p$ 
denotes the probability of applying each transform: Random gain ($p{=}0.25$, $\pm$6\,dB), 
additive background noise ($p{=}0.25$, signal-to-noise ratio (SNR) 0--25\,dB), pitch shift ($p{=}0.20$, $\pm$6 
semitones), colored noise ($p{=}0.10$, SNR 3--30\,dB).

\textbf{Overall Training Objective:} 
The proposed framework trains in a single progressive run 
(up to $\sim$\,1\,M steps). Phase transitions are triggered by validation-loss convergence: Phase 1 converges at $\sim$\,300\,k steps, after which the Phase 2 and Phase 3 objectives are introduced successively. The total loss $ \mathcal{L}_{\text{total}}$ is defined as:

\begin{equation}
    \mathcal{L}_{\text{total}} = 
    \underbrace{
        \overbrace{
            \underbrace{
                \strut \mathcal{L}_{\text{CE}}
            }_{\text{Phase 1}} 
            + \, \lambda_{\text{EL}} \cdot \mathcal{L}_{\text{L1}\text{--}\text{cos}}
        }^{\text{Phase 2}} 
        + \, \lambda_{\text{DTW}} \cdot \mathcal{L}_{\text{L1}(\text{DTW})}
    }_{\text{Phase 3}}
    \label{eq:total_loss}
\end{equation}

where $\lambda_{\text{EL}}$ and $\lambda_{\text{DTW}}$ are independently tunable scaling weights for the EL teacher and content-alignment losses. In our experiments, both were set to $0.6$. Note that within $\mathcal{L}_{\text{L1}\text{--}\text{cos}}$ (Eq.~\ref{eq:l1_cos}), the L1 and cosine components are 
equally weighted. The chosen weight factors down-weight the EL-related loss terms to reduce pathological domain overfitting, given the high HE/EL data imbalance. For training, we use Adam~\cite{kingma2015adam} with a linear warmup of 10\,k steps from $0$ to $10^{-4}$, followed by cosine annealing to $10^{-5}$ over $\sim$\,1\,M steps. The student encoder-only weights are exported for inference. Model training is conducted on a single NVIDIA A100 GPU and takes up to 72\,h.

\begin{table*}[t]
\centering
\begin{threeparttable}
\caption{Architectural summary of the student encoders. All models produce 64-dimensional
latent sequences at 50\,Hz. Conformer depthwise conv kernel $k{=}31$ (620\,ms receptive field);
CNN-Mamba causal conv kernel $k{=}4$.
$^{\ddagger}$Mamba has no hard context limit; the SSM hidden state is
fixed-size ($d_{\text{state}}{=}16$) but unbounded in temporal reach.}
\label{tab:architectures}
\small
\setlength{\tabcolsep}{4.1pt}
\begin{tabular}{@{}lllccccccccrrr@{}}
\toprule
\textbf{Model} & \textbf{Acoustic Frontend} & \textbf{Sequence Core} & \textbf{$d$} & \textbf{$L$} & \textbf{$H$} & \textbf{FFN} & \textbf{Act.} & \textbf{Conv\,$k$} & \textbf{Ctx} & \textbf{LA} & \textbf{Params} & \textbf{MB} \\
\midrule
CNN-Transformer  & SoundStream ($s{=}32$)~\cite{zeghidour2022soundstream} & Transformer (RoPE)~\cite{vaswani2017attention}  & 512  & 6   & 8   & $2{\times}$ & GELU & --  & 64                    & 1 & 17.36\,M & 69.4 \\
CNN-Conformer    & SoundStream ($s{=}32$) & Conformer (Macaron)~\cite{gulati20_interspeech} & 320  & 6   & 5   & $3{\times}$ & SiLU & 31  & 64                    & 1 & 16.51\,M & 66.0 \\
CNN-Mamba        & SoundStream ($s{=}32$) & Mamba (SSM)~\cite{gu2024mamba}         & 576  & 6   & --  & --          & --   & 4   & $\infty^{\ddagger}$   & 1 & 17.59\,M & 70.4 \\
\midrule
Mel-Conformer    & Mel-Spec.\ + CNN Pre-net  & Conformer (Macaron) & 384 & 6 & 8 & $4{\times}$ & SiLU & 31 & 64 & 1 & 21.94\,M & 87.8 \\
\bottomrule
\end{tabular}
\begin{tablenotes}    
\footnotesize
\item[] $d$: Model dimension;\enspace $L$: Number of layers;\enspace $H$: Attention heads;\enspace
FFN: Feed-forward expansion multiplier;\enspace Act.: FFN activation;\enspace
Conv\,$k$: Depthwise conv kernel;\enspace Ctx: Context window (frames);\enspace
LA: Per-layer lookahead (1\,frame\,=\,20\,ms);\enspace
MB: Footprint at FP32 precision;\enspace $-$: Not applicable.
\vspace{-0.30cm}
\end{tablenotes}              
\end{threeparttable}   
\end{table*}

\subsection{Student Architectures}
\label{sec:architectures}
\begin{figure}[t]
  \centering
    \includegraphics[width=1.0\linewidth]{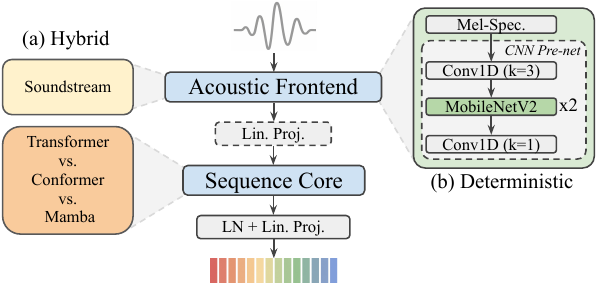}
    \vspace{-0.5cm}
  \caption{Generalized architecture of the student encoders. (a)~Hybrid variants (CNN-Transformer, CNN-Conformer, CNN-Mamba). (b)~The Mel-Conformer (Deterministic). Specific configurations for the shared frontend and the interchangeable sequence core are detailed in Table~\ref{tab:architectures}.}
  \label{fig:encoderarch}
  \vspace{-0.30cm}
\end{figure}

We evaluate three architecture families: fully CNN baselines, hybrid sequence models, and a deterministic mel-spectrogram frontend variant, all producing 64-dimensional latent representations at 50\,Hz. The latter two are detailed in Table~\ref{tab:architectures} and Fig.~\ref{fig:encoderarch}.


\textbf{Fully Convolutional Baselines:} 
AudioDec~\cite{wu2023audiodec} and SoundStream~\cite{zeghidour2022soundstream} encoders are evaluated as baselines since they serve as the standard content encoders in state-of-the-art streaming VC pipelines (e.g., StreamVC~\cite{yang2024streamvc}, RTVC~\cite{liu2025rtvc}, SynthVC~\cite{guo2025synthvc}). Both encoders rely entirely on stacked causal 1D-convolutions, but differ in their residual block configuration. We adopt them to downsample the raw waveform by $320\times$ (cascaded strides of 2, 4, 5, 8) to extract 50 Hz features. For both encoders, we use a base channel scale of 64. The FiLM conditioning layer from the original SoundStream encoder~\cite{zeghidour2022soundstream} is omitted. Encoders from other EL VC systems are precluded by the absence of open-source implementations~\cite{Lester2024, MaDing2025}.

\textbf{Hybrid Sequence Models:} 
To enable low-latency streaming, all presented encoder architectures employ causal operations. To isolate the impact of the sequence modeling core, our three hybrid architectures (Transformer~\cite{vaswani2017attention}, Conformer~\cite{gulati20_interspeech}, Mamba~\cite{gu2024mamba}) share the identical SoundStream encoder frontend for downsampling (Fig.~\ref{fig:encoderarch}\,(a)), with its base channel scale (s) reduced from 64 to 32 to accommodate the sequence models' parameter footprints. We 
constrain their parameter budget to match the convolutional baselines (AudioDec, SoundStream: $\sim$\,18.55\,M; Hybrids: 16.5--17.6\,M). All attention-based variants (CNN-Transformer and CNN-Conformer) use causal self-attention with Rotary Position Embeddings (RoPE). CNN-Mamba uses no explicit positional encoding. The CNN-Conformer injects causal depthwise convolutions (Macaron-style) to better capture local acoustic textures. The CNN-Mamba replaces quadratic attention with a Selective State Space Model (SSM)\footnote{https://github.com/state-spaces/mamba}, offering linear time complexity and a fixed-size recurrent hidden state ($d_{\text{state}}{=}16$) that replaces the KV-cache of attention-based models. For these attention-based architectures, a causal context window of 64 frames (1.28\,s) is enforced to bound streaming memory. All hybrid architectures apply a one-frame (20\,ms) lookahead per layer, compounding to a 120\,ms effective streaming horizon across the six-layer stack. At inference, this horizon is realized by frame-synchronous incremental decoding with per-layer output hold-back. Each layer releases the representation of frame $t$ only after the key and value of its single lookahead frame $t{+}1$ have been computed and cached. The deferral composes additively across the six layers, so every emitted frame observes exactly the bounded future context seen during training. Rotary-embedded queries are retained in the cache to avoid recomputation. The specific hyperparameters of each configuration (Table~\ref{tab:architectures}) were selected through a small-scale empirical search.

\textbf{Deterministic Acoustic Baseline:} 
The Mel-Conformer (21.9\,M) departs from the hybrid design by replacing the 
learned waveform frontend with a fixed mel-spectrogram ($n_\text{fft}{=}1024$, $\text{hop}{=}320$, 80 bins), preserving the 50\,Hz frame rate without trained downsampling (Fig.~\ref{fig:encoderarch}\,(b)). This eliminates the 
learned-parameter inference cost of the acoustic frontend and redirects the 
additional parameter capacity into the sequence core. A lightweight CNN pre-net projects the mel features into the sequence core: CausalConv1d ($k{=}3$, stride-1) expands $80{\to}d$, followed 
by two MobileNetV2-style~\cite{sandler2018mobilenetv2} inverted residual 
blocks (pointwise Conv1d ($k{=}1$) expand ${\times}2$ $\to$ causal depthwise 
Conv1d ($k{=}3$) $\to$ pointwise Conv1d ($k{=}1$) project, with residual skip 
connection) and a final pointwise Conv1d ($k{=}1$). All operations are 
strictly causal. Normalization uses LayerNorm with SiLU activations throughout. Since the CNN pre-net directly emits $d$-dimensional features, the Mel-Conformer omits the linear input projection used by the hybrid variants.

\section{Experiments}
\label{sec:experiments}

\textbf{Teacher Ablation:} 
To select the optimal HE teacher and its most informative layer, we conduct a 
layer-wise probing study across seven frozen SSL teachers: HuBERT (base), 
HuBERT (large)~\cite{huber2021}, mHuBERT-147~\cite{zanonboito2024mhubert}, WavLM (base), 
WavLM (large)~\cite{wavlm2022}, WavLabLM-MS40~\cite{chen2023joint}, and W2v-BERT~2.0~\cite{barrault2023seamless}. All model weights are kept frozen, 
only a lightweight CTC probe head is trained per layer. The probe consists of a LayerNorm, 
a linear projection to 512 dimensions, a two-layer bidirectional Long Short-Term Memory (BiLSTM), and a character-level CTC output layer. None of the evaluated models were exposed to either the HE speech or EL 
speech during pre-training, making this a zero-shot assessment of cross-domain 
representational transfer.

Each teacher is evaluated independently on both HE and EL speech, using an 80/10/10 
train/val/test split of the ELHE corpus (approximately 370 utterances per validation 
and test set). All splits are utterance-level. The CTC probe therefore observes every speaker, and WER/CER are read as a relative comparison across models. To control 
for split-specific bias, we run two cross-validated experiments with swapped val/test 
assignments (val$\leftrightarrow$test). Per-layer probes are trained until convergence
(AdamW~\cite{loshchilov2019decoupled}, $\text{lr}{=}10^{-3}$, cosine annealing with 1-epoch warmup to 
$5{\times}10^{-4}$). The checkpoint with lowest 
validation WER is selected. For the two best-performing SSL teachers, we additionally run 
four random seed replicates to confirm result 
stability. The reported WER/CER are computed over the pooled test predictions across seeds and splits, with 95\% confidence intervals (CI) from a 1{,}000-resample utterance-level bootstrap.

\textbf{Student Encoder Ablation:}
Unlike the teacher ablation, which probes intermediate layers of high-dimensional SSL encoders (typically 768--1280 dimensions), we evaluate only the final output of each student encoder (64 dimensions), the representation directly used in downstream tasks (e.g.\ VC, ASR), with the CTC probe's input projection adapted accordingly. We note that this bottleneck may inherently disadvantage the student probe relative to the teacher probe, as the student must compress all phonetic content into a substantially lower-dimensional space. To improve comparability 
across architectures, each student is probed twice with validation and test sets swapped (val$\leftrightarrow$test) again, we report the mean WER and CER across both runs.

 \textbf{Feature Space Analysis:}
  To assess whether the training objective improves EL--HE domain alignment in the
  learned representation, we extract frame-level features from the output of the CNN-Transformer student encoder for 8 EL utterances across 4
  speakers (2 per speaker) and their matched HE counterparts of identical lexical
  content. Silero voice activity detection (VAD)\footnote{https://github.com/snakers4/silero-vad}
 is used to exclude non-speech frames prior to extraction.
  For visualization, we apply Uniform Manifold Approximation and Projection (UMAP)~\cite{mcinnes2018umap} ($n_\text{neighbors}{=}15$, $\text{min\_dist}{=}0.1$, fitted
  independently per condition). Quantitative metrics are computed in the original 64-dimensional space on features standardized per phase by a scaler fit on the pooled EL and HE frames of that training phase. We report two
  projection-free metrics. The average paired EL--HE centroid distance (lower is
  better) and the 5-fold cross-validated accuracy of a logistic-regression domain
  classifier trained to separate EL from HE frames (0.5 means the domains are indistinguishable
  domains, i.e.\ better alignment).

\textbf{Real-Time Factor (RTF) \& Latency:}
We benchmark each encoder on a single CPU core
(AMD Ryzen AI~7~PRO~360; 1 thread; PyTorch~2.10, FP32) in stateful
mode. Convolutional baselines maintain CausalConv1d ring
buffers, hybrid models use incremental KV-cache or SSM state updates.
A 200-frame warmup is discarded, state is preserved into the
1{,}000-frame measurement window. RTF is mean per-chunk inference
time divided by chunk duration. RTF\,$<$\,1.0 denotes real-time.

Since latency is determined by the architectural hold-back, we benchmark in the frame-synchronous regime (i.e., 20 ms frame per call). The total algorithmic latency is $(1{+}L_h)\times 20$ ms, where $L_h$ is the cascaded hold-back: $0$ frames for the strictly-causal convolutional baselines and $6{=}6{\times}1$ frames for the six-layer hybrid encoders, yielding 20 ms and 140 ms respectively. We verified that this streaming schedule is numerically equivalent to offline processing of the full utterance (mean cosine distance below $5\times10^{-5}$ across all encoders), confirming that the offline WER of Table~\ref{tab:student_ablation} transfers without degradation to streaming.

ONNX Runtime (ORT, opset~18, 1~thread) is reported at 20\,ms. All model states, CNN ring buffers, KV/conv caches, SSM states, and
RoPE position offsets, is threaded as input/output tensors,
position indices are resolved via a Gather operation to ensure
dynamic lookup across calls. CNN-Mamba is exported as a per-frame stateful
step graph, avoiding the sequential scan loop that precludes full
model tracing.

\begin{figure}[t]
  \centering
    \includegraphics[width=1.0\linewidth]{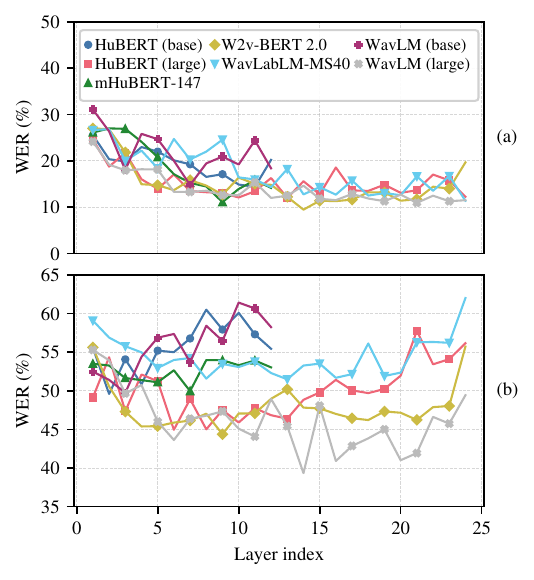}
    \vspace{-0.5cm}
  \caption{Layer-wise probing results (WER) for the evaluated SSL teacher models on (a) HE data and (b) EL data.}
  \label{fig:teacherablation}
  \vspace{-0.30cm}
\end{figure}

\section{Results \& Discussion}
\label{sec:results}

\textbf{Teacher Ablation:} 
Layer-wise probing reveals a great contrast between evaluation domains.
On HE speech (Fig.~\ref{fig:teacherablation}\,(a)), WER generally decreases
across intermediate layers, with no model achieving its lowest WER in the
final layer.
W2v-BERT~2.0 achieves the lowest HE WER at 9.5\% (layer~14),
while WavLM (large) (layer 21) and mHuBERT-147 (layer 9) follow both with approximately 11\%.
On EL speech (Fig.~\ref{fig:teacherablation}\,(b)), layer-wise performance is
largely stationary and noisy across all models, indicating severe domain
mismatch that limits feature extraction.
HuBERT (base) and HuBERT (large) exhibit monotonic WER degradation with
depth, showing that deeper layers increasingly over-specialize to
the pre-training domain.
WavLM (large) (layer 14) achieves the lowest EL WER at 39.3\%; W2v-BERT~2.0 (layer 9)
and HuBERT (large) (layer 6) follow with approximately 44\%, while the
remaining four models cluster near 50\%.
WavLM (large)'s relative robustness is likely attributable to its noise-gated
pre-training objective~\cite{wavlm2022}.

Despite W2v-BERT~2.0 achieving the lowest HE WER, initial 
experiments showed that the student trained with mHuBERT-147 as the HE teacher achieved better downstream performance.
We attribute this to a known capacity-gap effect in knowledge
distillation~\cite{cho2019efficacy}, wherein a substantially larger teacher
produces targets that are harder for a compact student to approximate
reliably. The mHuBERT-147 model (layer~10) is therefore selected as the HE teacher.

\begin{table}[h]
\centering
\caption{Downstream ASR probing WER\,/\,CER (\%) for the top-3 teachers
(zero-shot, best layer) and distilled student encoders under
progressively richer training objectives.
Subscripts: $\pm$\,95\% CI.}
\label{tab:student_ablation}
\begin{threeparttable}
\small
\setlength{\tabcolsep}{3pt}
\begin{tabular}{@{}lcccc@{}}
\toprule
\multirow{2}{*}{\textbf{Model}} &
\multicolumn{2}{c}{\textbf{HE}} &
\multicolumn{2}{c}{\textbf{EL}} \\
\cmidrule(lr){2-3}\cmidrule(lr){4-5}
& WER$\downarrow$ & CER$\downarrow$ & WER$\downarrow$ & CER$\downarrow$ \\
\midrule
\multicolumn{5}{l}{\textit{SSL Teachers (zero-shot):}} \\
\quad W2v-BERT~2.0   & $\mathbf{9.5}_{(\pm0.6)}$           & $\mathbf{2.1}_{(\pm0.1)}$  & $44.3_{(\pm1.3)}$  & $14.1_{(\pm0.5)}$  \\
\quad WavLM (large)    & $10.9_{(\pm1.5)}$          & $2.5_{(\pm0.4)}$           & $\mathbf{39.3}_{(\pm2.8)}$  & $\mathbf{12.7}_{(\pm1.1)}$  \\
\quad mHuBERT-147$^{*}$    & $11.1_{(\pm0.6)}$          & $2.5_{(\pm0.2)}$           & $49.9_{(\pm1.4)}$  & $18.0_{(\pm0.6)}$  \\
\midrule
\multicolumn{5}{l}{\textit{Causal Student (Phase 1):}} \\
\quad CNN-Transformer  & $17.2_{(\pm2.0)}$ & $\mathbf{3.9}_{(\pm0.6)}$ & $68.1_{(\pm3.4)}$ & $26.8_{(\pm1.8)}$ \\
\midrule
\multicolumn{5}{l}{\textit{Causal Student (Phase 2):}} \\
\quad CNN-Transformer  & $17.4_{(\pm1.9)}$ & $4.1_{(\pm0.6)}$ & $21.2_{(\pm3.0)}$ & $8.4_{(\pm1.4)}$ \\
\midrule
\multicolumn{5}{l}{\textit{Causal Students (Phase 3):}} \\
\quad CNN-Transformer
    & $17.3_{(\pm1.9)}$ & $4.0_{(\pm0.6)}$
    & $21.7_{(\pm3.0)}$ & $8.5_{(\pm1.4)}$ \\
\quad CNN-Conformer
    & $17.4_{(\pm1.9)}$ & $4.1_{(\pm0.6)}$
    & $21.8_{(\pm3.1)}$ & $8.4_{(\pm1.4)}$ \\
\quad CNN-Mamba      & $\mathbf{16.2}_{(\pm1.9)}$ & $\mathbf{3.9}_{(\pm0.6)}$ & $22.5_{(\pm3.1)}$ & $8.8_{(\pm1.5)}$ \\
\quad Mel-Conformer  & $17.2_{(\pm2.0)}$ & $\mathbf{3.9}_{(\pm0.6)}$ & $\mathbf{21.2}_{(\pm3.0)}$ & $\mathbf{8.3}_{(\pm1.4)}$ \\
\midrule
\multicolumn{5}{l}{\textit{Convolutional Students (Phase 3):}} \\
\quad AudioDec          & $24.8_{(\pm2.2)}$ & $5.9_{(\pm0.7)}$ & $41.3_{(\pm3.4)}$ & $15.1_{(\pm1.7)}$ \\
\quad SoundStream & $22.6_{(\pm2.2)}$ & $5.3_{(\pm0.7)}$ & $32.3_{(\pm3.6)}$ & $12.8_{(\pm1.8)}$ \\
\bottomrule
\end{tabular}
\begin{tablenotes}              
\footnotesize
\item[$*$] Fine-tuned mHuBERT-147-BNF-CTC on HE+EL jointly and evaluated via direct CTC decode through a 100-dim bottleneck (not probe, not directly comparable): EL WER 16.6\% and HE WER 13.6\%.
\end{tablenotes}              
\end{threeparttable}          
\vspace{-0.3cm}
\end{table}

\begin{figure*}[t]
  \centering
    \includegraphics[width=1.0\linewidth]{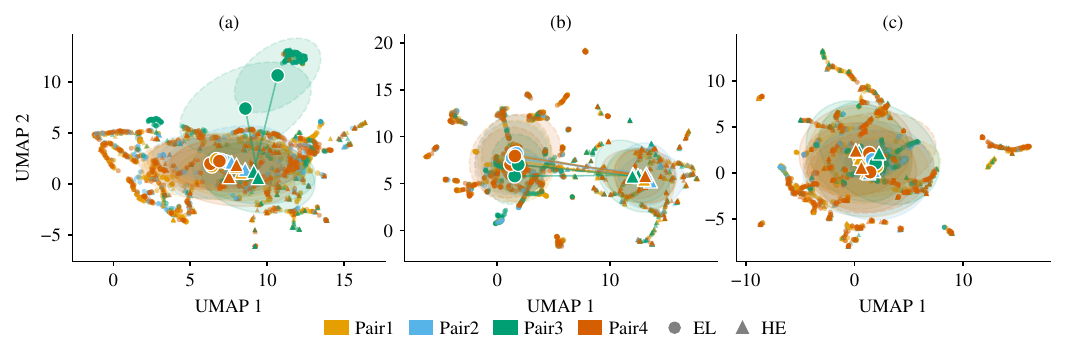}
    \vspace{-0.9cm}
  \caption{UMAP projections of the CNN-Transformer student encoder output (64-dim).
Color: Speaker pair; marker: Modality ({\large$\bullet$}~EL, $\blacktriangle$~HE);
lines connect matched EL--HE centroids.
(a)~Phase~1, (b)~Phase~2, (c)~Phase~3.
Each panel is fitted independently.}
  \label{fig:umap}
  \vspace{-0.35cm}
\end{figure*}

\textbf{Student Ablation:}
Table~\ref{tab:student_ablation} reports downstream ASR probing results. The Phase~1 to Phase~2 transition validates the dual-teacher design.
Introducing EL bottleneck regression drops CNN-Transformer EL WER from
68.1\% to 21.2\% while leaving HE WER nearly unchanged (17.2\% to 17.4\%).
The addition of the DTW alignment loss in Phase 3 leaves CNN-Transformer EL WER statistically unchanged (21.2\%--21.7\%, within CI). Its primary effect is geometric, as confirmed by the Feature Space Analysis below (Fig.~\ref{fig:umap}). The fully convolutional baselines (AudioDec \& SoundStream) underperform all distilled
students under identical training conditions
(AudioDec: EL WER 41.3\%, SoundStream: EL WER 32.3\%), indicating that a sequence
modeling core meaningfully complements distillation; however, this comparison is
confounded by lookahead, as AudioDec/SoundStream operate with zero lookahead by
construction, unlike the 120\,ms shared by all hybrid students, so the gap cannot be
attributed to the sequence core alone. Among the other students, CNN-Transformer and CNN-Conformer yield
similar EL WER (21.7\%--21.8\%).
CNN-Mamba achieves the best HE WER overall (16.2\%), but trails on EL WER (22.5\%). The difference to the best EL student falls within overlapping CI. The Mel-Conformer achieves the lowest EL WER/CER across all students (WER: 21.2\%, CER: 8.3\%), representing a $\sim$\,46\% relative EL WER reduction over the strongest SSL baseline, WavLM (large) with 39.3\%.
Its advantage traces to the mel-spectrogram frontend: Eliminating the
learned downsampling stack reduces per-call inference cost and
allocates the saved parameter budget to a larger sequence core.

The remaining gap between the best student (EL WER 21.2\%)
and the pathological teacher's direct CTC decode (EL WER 16.6\%;
footnote~$*$) is expected. The teacher operates non-causally on full-utterance context with a CTC head jointly optimized alongside its encoder, whereas the student is constrained to a 120\,ms causal lookahead, and a 64-dimensional output bottleneck.

\textbf{Feature Space Analysis:}
Fig.~\ref{fig:umap} shows UMAP projections of the student encoder output under
the three training conditions. In the single-teacher~(Phase 1) case~(a), EL and HE
features lie in a moderately separated region (mean centroid distance 4.20,
domain-classifier accuracy 0.77), but this does not reflect meaningful
phonetic alignment. The corresponding EL WER reaches 68.1\%, indicating
representation collapse in which both modalities map to an uninformative region
regardless of phonetic content. Notably, speaker pair~3, (representing a
particularly severe EL speaker) appears as two visually isolated clusters,
inflating the mean centroid distance at this stage.
The two-teacher~(Phase 2) objective~(b) resolves this
collapse and yields distinct, internally coherent per-utterance clusters,
consistent with the recovered phonetic structure (EL WER 21.2\%). The two
domains, become more separable (mean centroid distance 4.13,
classifier accuracy 0.99): Learning from two teachers drives EL and HE
into distinct sub-clusters. Adding the $\mathcal{L}_{\text{L1}(\text{DTW})}$ alignment
loss~(Phase 3) in~(c) pulls matched EL--HE pairs together (mean centroid distance 1.34)
and renders the domains substantially harder to separate
(accuracy 0.65) while preserving inter-utterance
separability. Even the isolated speaker pair~3 converges into
the joint cluster structure, suggesting the alignment objective is robust to severe EL acoustic distortion.

\textbf{Real-Time Factor \& Latency:}
Results are reported in Table~\ref{tab:rtf_updated}. Streaming is performed
frame-synchronously (i.e., 20\,ms frame per call), the evaluated operating point,
at which all four hybrid encoders are real-time (RTF\,$<$\,1.0) whereas the
strictly-causal convolutional baselines are not (RTF\,$>$\,1.41), being
dominated by per-call overhead across their stacked dilated convolutions. The
Mel-Conformer is the fastest model (RTF of \,0.64), as its fixed
mel-spectrogram frontend replaces the learned waveform-downsampling stack.
Processing six frames per call (120\,ms column) amortizes this per-call
overhead and brings every model below
an RTF of 0.65. Export to
ONNX Runtime yields a further 2.1--2.4$\times$ speedup for all encoders through operator fusion. CNN-Mamba reaches an RTF of 0.44, as its sequential selective scan does not benefit from frame batching (Section~\ref{sec:experiments}). These throughput gains do not reduce algorithmic latency, which remains fixed by the per-layer hold-back at 20\,ms (convolutional) and 140\,ms (hybrid) (Table~\ref{tab:rtf_updated}). These latencies are below the low-latency recommendations for real-time telecommunication applications~\cite{itu2003g114}.

In summary, the Mel-Conformer yields the best EL accuracy and lowest RTF of all evaluated architectures (Table~\ref{tab:rtf_updated}), while 
achieving HE accuracy competitive with the other students within 
overlapping CI. This confirms the hypothesis that the fixed mel-spectrogram frontend contributes in two ways: it redirects parameter budget toward a larger Conformer sequence core, improving representational quality, while eliminating the
per-call waveform-downsampling cost, reducing RTF, making the Mel-Conformer optimal across all evaluated architectures on EL accuracy and efficiency.

\begin{table}[h]
\centering
\caption{Single-threaded CPU streaming benchmark.
PyTorch RTF at 20\,ms (frame-synchronous) and 120\,ms (six-frame) chunk.
ONNX Runtime (ORT, opset~18) at 20\,ms.
Algorithmic latency is fixed by per-layer hold-back.}
\label{tab:rtf_updated}
\small
\begin{threeparttable}
\setlength{\tabcolsep}{6pt}
\begin{tabular}{@{}l cc c c@{}}
\toprule
\multirow{2}{*}{\textbf{Model}} &
\multicolumn{2}{c}{\textbf{PyTorch RTF}} &
\textbf{ONNX RTF} &
\textbf{Alg.\ Lat.} \\
\cmidrule(lr){2-3} \cmidrule(lr){4-4} \cmidrule(lr){5-5}
& 20\,ms & 120\,ms & 20\,ms & (ms) \\
\midrule
\multicolumn{5}{l}{\textit{Baselines}} \\
\quad SoundStream & 1.43 & 0.62 & 0.72 & 20 \\
\quad AudioDec    & 1.42 & 0.64 & 0.70 & 20 \\
\midrule
\multicolumn{5}{l}{\textit{Hybrid sequence models}} \\
\quad CNN-Transformer   & 0.87 & 0.47 & 0.37 & 140 \\
\quad CNN-Conformer     & 0.95 & 0.56 & 0.40 & 140 \\
\quad CNN-Mamba\tnote{$\dagger$} & 0.87 & 0.48 & 0.44 & 140 \\
\midrule
\multicolumn{5}{l}{\textit{Deterministic frontend}} \\
\quad Mel-Conformer & \textbf{0.64} & \textbf{0.25} & \textbf{0.30} & 140 \\
\bottomrule
\end{tabular}
\begin{tablenotes}
\footnotesize
\item[$\dagger$] ONNX uses per-frame stateful-step export. PyTorch uses the sequential SSM fallback (mamba-ssm CUDA kernel
  unavailable on CPU).
\end{tablenotes}
\end{threeparttable}
\vspace{-0.35cm}
\end{table}

\section{Limitations \& Future Work}
\label{sec:futurework}
While the proposed framework demonstrates strong results, some limitations point toward directions for future work. Phase~3 requires parallel EL--HE recordings, restricting applicability
to speakers for whom such paired data exists. Future work will
investigate pseudo-parallel and speaker-adaptive alignment strategies
to relax this constraint.
The present evaluation relies exclusively on downstream ASR probing,
which does not capture prosodic structure or synthesis quality necessary for VC deployment. End-to-end VC evaluation with perceptual metrics is deferred to a
follow-up study, including additional prosody modules for enhanced pathological to HE speech rehabilitation.
RTF results are specific to the evaluated CPU platform. We aim to investigate embedded and neural processing unit (NPU) targets as well as quantization techniques for optimized deployment.

\section{Conclusion}
\label{sec:conclusion}

We presented a multi-teacher knowledge distillation framework for training
a lightweight, streaming-compatible content encoder that generalizes across
HE and EL speech.
By progressively distilling discrete phonetic targets from a frozen SSL model and
continuous bottleneck features from an EL-fine-tuned ASR model, and
progressively introducing a DTW-based domain alignment objective, the
proposed three-phase training effectively bridges the EL--HE domain gap
without catastrophic forgetting of HE-domain representations.
The best student, a Mel-Conformer encoder whose fixed mel-spectrogram
frontend eliminates learned waveform downsampling in favor of a larger
sequence core, achieves a 46\% relative EL WER reduction over
the strongest zero-shot SSL baseline (39.3\% to 21.2\%) while operating at
an RTF of 0.30 (ONNX Runtime) on a single CPU core with 21.9\,M parameters.
Feature space analysis confirms that the alignment objective geometrically
maps the EL--HE latent space without degrading phonetic discriminability.
Future work will integrate the encoder into a full end-to-end streaming VC
pipeline with prosody modeling, and extend evaluation to perceptual
naturalness and intelligibility metrics on live EL speech.

\section*{ACKNOWLEDGMENTS}
This research was funded in part by the Austrian Science Fund (FWF) [10.55776/PAT5948223] and utilized the Austrian Scientific Computing (ASC) infrastructure. Gemini (Google) was used for background literature research and manuscript proofreading. Claude Code (Anthropic) was used in the development and implementation of the codebase.


\bibliographystyle{IEEEtran}
\bibliography{refs}

\end{document}